\documentclass[twocolumn,amsmath,amssymb,superscriptaddress]{revtex4}

\usepackage{graphicx}
\usepackage{dcolumn}
\usepackage{bm}
\usepackage{amsmath}
\usepackage{float}

\begin{document}

\title{Nuclear shape transitions and elastic magnetic electron scattering}

\author{B.~Hern\'andez}
\affiliation{
Instituto de Estructura de la Materia, IEM-CSIC, Serrano
123, E-28006 Madrid, Spain}
\affiliation{
  Departamento de Estructura de la Materia, F\'{\i}sica T\'ermica y Electr\'onica,
  and IPARCOS, Facultad de Ciencias F\'\i sicas, Universidad Complutense de Madrid,
  Madrid E-28040, Spain}

\author{P.~Sarriguren}
\email{p.sarriguren@csic.es}
\affiliation{
Instituto de Estructura de la Materia, IEM-CSIC, Serrano 123, E-28006 Madrid, Spain}

\author{O.~Moreno}
\author{E.~Moya~de~Guerra}
\affiliation{
  Departamento de Estructura de la Materia, F\'{\i}sica T\'ermica y Electr\'onica,
  and IPARCOS, Facultad de Ciencias F\'\i sicas, Universidad Complutense de Madrid, 
  Madrid E-28040, Spain}

\author{D.~N.~Kadrev}
\author{A.~N.~Antonov}
\affiliation{
Institute for Nuclear Research and Nuclear Energy,
Bulgarian Academy of Sciences, Sofia 1784, Bulgaria
}

\date{\today}

\begin{abstract}

  Backward elastic electron scattering from odd-$A$ nuclear targets is characterized
  by magnetic form factors containing precise information on the nuclear structure.
  We study the sensitivity of the magnetic form factors to structural effects related
  to the evolution and shape transitions in both isotopic and isotonic chains.
  Calculations of magnetic form factors are performed in the plane-wave Born
  approximation. The nuclear structure is obtained from a deformed self-consistent
  mean-field calculation based on a Skyrme HF+BCS formalism. Collective effects are
  included in the cranking approximation, whereas nucleon-nucleon correlations are
  taken into account in the coherent density fluctuation model. The evolution of the
  magnetic form factors is found to exhibit signatures of shape transitions that show
  up in selected isotopic and isotonic chains involving both stable and unstable nuclei.
  Several cases are identified as suitable candidates for showing such fingerprints of
  shape transitions. A new generation of electron scattering experiments involving
  electron-radioactive beam colliders will be available in the near future, leading
  to a renewed interest in this field.

\end{abstract}

\maketitle

\section{Introduction}

Electron scattering from nuclei has been well established in the past as a tool to
provide detailed information on nuclear charge and current distributions
\cite{hofstadter,forest,donnwal,heisenberg}. It is well known that the weakness of the
electromagnetic interaction relative to the nuclear force, as well as its accurate
quantum electrodynamics description, result in electron scattering being a suitable
and powerful tool to study nuclear charge distributions and radii, transition
probabilities, momentum distributions, and spectroscopic factors from elastic,
inelastic, and quasielastic channels \cite{vri87,frullani,udias}. 

Electron scattering can be also used as an additional tool to complement the information
obtained from other probes. This could be the case of magnetic dipole ($M1$)
excitations in nuclei, where it is well known  that complementary
information is obtained when using different electromagnetic
$(\gamma , \gamma '),\, (e,e')$ or hadronic $(p,p')$ probes \cite{heyde,sarri_m1}.
Electron scattering may also be used as a tool to investigate weak processes,
such as parity-violating electron
scattering \cite{don_parity,moreno_parity,prex} or other processes involving
neutrinos \cite{langanke,superscaling}.

Although charge scattering is in most cases the dominant contribution to the cross
section, electrons also interact with the nuclear electromagnetic current distributions,
whose contributions can be isolated with proper choices of the kinematical conditions
of the process. The experimental observation of the electric and magnetic
form factors provides information on the convection and magnetization currents within
the nucleus. In particular, elastic magnetic electron scattering provides fine details
of the nuclear ground-state current and magnetization distributions
\cite{donnelly,elvira,annals_elvira}.

Contrary to the case of charge scattering, where all the protons contribute coherently
to the nuclear response, the magnetic scattering response in odd-$A$ nuclei depends to
a large extent on the single-particle properties of the valence nucleon wave function,
whereas the collective aspects give minor contributions in most cases, but may play an
important role in modulating the single-particle contributions. In addition,
magnetic electron scattering may provide information on both protons and neutrons on an
equal footing, since their intrinsic magnetic moments are similar in magnitude.
Similarly to the case of charge scattering, magnetic scattering has been studied
in stable nuclei from different theoretical frameworks \cite{donnelly}, including shell
model \cite{radhi_03,radhi_07,jassim,samma}, relativistic mean field \cite{wang}, and
deformed mean field models \cite{elvira,diep,kova,graca,ps,berdi,nk,merino,yao,wang_20}.

In a previous work \cite{merino}, we showed that the plane-wave Born approximation
(PWBA) together with a self-consistent deformed Skyrme HF+BCS formalism for the nuclear
structure calculation are able to reproduce the experimental form factors of both
spherical and deformed stable nuclei. Therefore, we have at hand a reliable procedure
to address the study of unstable nuclei and to predict the magnetic
form factors of nuclei not yet measured. Based on this method, we plan in this work to
study the sensitivity of the magnetic form factors against structural changes of the
nucleus. The best scenario for this study involves chains of odd-$A$ isotopes with odd
$Z$ or chains of odd-$A$ isotones with odd $N$, whose ground-state spin/parity ($I^\pi$),
determined by the unpaired nucleon, only changes if the structure of the nucleus
is changed. Therefore, we look first for experimental fingerprints of structural
changes and identify some candidates where this study could be of particular interest.
The structural change will be related to a shape transition along the
chain \cite{heyde_shape,rayner}. 

The study of isotopic or isotonic chains involves in general unstable nuclei. One
criterion for our selection of the chains to be studied is based on the eventual
feasibility of electron scattering experiments from the nuclei in the chain.
Therefore, we focus on the unstable nuclei that are close to stability, in addition
to the stable ones.
The experimental difficulties found when dealing with unstable targets have limited
the present knowledge to stable nuclei, but it is expected that the new generation of
electron-ion colliders at radioactive nuclear beam facilities \cite{sudasimon} will
overcome some of the technical difficulties. This is the case of ELISe
(FAIR-GSI) \cite{elise} and SCRIT (RIKEN) \cite{scrit}. The conceptual design and the
scientific challenges of the electron-ion collider ELISe (Electron-Ion Scattering in
a Storage Ring) can be found in Ref. \cite{anton_nima}. In SCRIT (Self-Confining
Radioactive Ion Target) a circulating beam of electrons scatters off ions stored in
a trap \cite{scrit_npn,ohnishi}. In this facility, an elastic electron scattering
experiment on $^{132}$Xe has already been performed \cite{scrit_xe}, demonstrating its
feasibility.
Electron scattering on unstable nuclei could be a tool to study the evolution of the
charge distributions in isotopic chains and therefore, to test theoretical models
aiming to predict nuclear charge distributions. Examples of these studies can be
found in Refs. \cite{garrido,antonov_ff,dong,roca_13,radhi_18,liang_18}.

The structure of this article is the following. In section \ref{form} we present the
theoretical framework and basic expressions to calculate the magnetic form factors in
deformed nuclei. We also introduce the spherical limit of these calculations, as well
as the effects of short-range correlations in this limit. Section III contains the
results obtained. The latter are given, first for static moments and then for the
magnetic form factors in the isotopic and isotonic chains selected. Section IV
contains the conclusions of the work.

%%%%%%%%%%%%%%%%%%%%%%%%%%%%%%%%%%%%%%%%%%%%%%%%%%%%%%%%

\section{Theoretical formalism}
\label{form}

\subsection{Electron scattering form factors}

The basic aspects of the formalism of electron scattering from deformed nuclei that we
follow in this work was introduced earlier \cite{elvira,kova}. The work in \cite{kova}
demonstrated for the first time that deformation has to be included in the theoretical
description in order to reproduce the experimental data for the deformed nucleus
$^{181}$Ta. Since then, the method has been applied to different
cases \cite{merino,graca,ps,berdi,nk}, where the
sensitivities of the results to different approximations concerning nuclear structure
and reaction mechanism were studied. In particular, it has been shown that the magnetic
form factors of deformed nuclei may differ considerably from those of spherical nuclei
\cite{merino}. Here, we briefly summarize this formalism. Following the notation of
Ref. \cite{elvira}, the general cross section for ultrarelativistic electron scattering
for transitions from the nuclear ground state ($I_i$) to final states ($I_f$) in PWBA
and for unpolarized projectiles and unoriented targets, is given by

\begin{equation}
\left. \frac{d\sigma}{d\Omega} \right| _{I_i\rightarrow I_f} = 4\pi \sigma_M f_{\rm rec}^{-1}
\left[ V_L |F_L|^2 + V_T |F_T|^2 \right] \, ,
\label{eq1}
\end{equation}
in terms of the Mott cross section

\begin{equation}
\sigma_M = \left[\frac{\alpha \cos(\theta/2)}{2\epsilon_i \sin^2 (\theta/2)}\right]^2 \, ,
\end{equation}
and a recoil factor $f_{\rm rec}$. The cross section is separated into longitudinal (L)
and transverse (T) parts, weighted with different kinematical factors.
%\begin{equation}
%f_{\rm rec} = 1+\frac{2\epsilon_i \sin^2(\theta/2)}{M_{\rm Target}}\, .
%\end{equation}
The dependence on the electron kinematics is given by the L and T
Rosenbluth factors,

\begin{equation}
V_L= (Q^2/q^2)^2\, , \quad V_T = \tan^2(\theta /2)-(Q^2/2q^2) \, ,
\label{eq4}
\end{equation}
where the kinematical variables are defined so that an incident electron with 
four-momentum $k_{i\mu}=(\epsilon_i,{\bm k_i})$ is scattered from the nucleus through an
angle $\theta$ to four-momentum $k_{f\mu}=(\epsilon_f,{\bm k_f})$ by exchanging a virtual
photon with four-momentum $Q=(\omega,{\bm q})$. 

Whereas the longitudinal form factors receive coherent contributions from all the
charged nucleons, the transverse  form factors are basically single-particle observables
that depend mostly on the properties of the unpaired nucleon in the outermost shell.
Hence, the longitudinal contribution (charge scattering) is dominant in most cases
and the transverse contribution can only be disentangled using special kinematical
conditions. In particular, backward scattering ($\theta = 180^{\circ} $) is commonly
used to measure $F_T$.

The dependence on the nuclear structure is contained in the $q$-dependent longitudinal
and transverse form factors, which are written in terms of Coulomb (C), transverse
electric (E), and transverse magnetic (M) multipoles,

\begin{equation}
|F_L|^2 = \sum _{\lambda \ge 0} |F^{C\lambda}|^2\, , \quad
|F_T|^2 = \sum _{\lambda \ge 1} \left[ |F^{M\lambda}|^2 + |F^{E\lambda}|^2 \right] ,
\end{equation}
which are defined as the reduced matrix elements of the multipole operators  
$\hat{T}^{\sigma\lambda}$ between initial and final nuclear states

\begin{equation}
|F^{\sigma\lambda}|^2 = \frac{ \left| \langle I_f || \hat{T}^{\sigma\lambda}(q)||I_i
\rangle \right|^2} {2I_i+1} \, .
\label{f_rme} 
\end{equation}

For elastic scattering, parity and time reversal invariance imply that only even 
Coulomb and odd transverse magnetic multipoles contribute. Then, at $\theta=180^{\circ}$
only odd magnetic multipoles survive in PWBA,

\begin{equation}
| F_T(q) |^2 = \sum_{\lambda={\rm odd}} | F^{M\lambda} |^2 \, .
\label{ftfm}
\end{equation}

The magnetic multipole operators are defined as 

\begin{equation}
\hat{T}^{M\lambda}_{\mu}(q) = i^{\lambda}  \int d{\bf r}\, j_{\lambda}(qr) 
{\bf Y}^{\mu}_{\lambda\lambda}(\Omega_r) \cdot \hat{\bf{J}}  ({\bf r}) \, ,
\label{tensor}
\end{equation}
where $\hat{\bf{J}} ({\bf r})$ is the current density operator that contains both
convection and magnetization components arising from the motion and from the intrinsic
magnetic moments of the nucleons, respectively.

Center of mass and finite nucleon size corrections are included in the calculations.
For the center of mass correction we use the usual factor obtained in the
harmonic-oscillator approximation, $f_{\rm c.m.}=\exp{\left( q^2A^{-2/3}/4 \right) }$.
In the magnetization currents, we use bare nucleon magnetic moments,
$\mu_{s}^p=2.793\ \mu_N$, $\mu_{s}^n=-1.913\ \mu_N$, corrected with dipole form 
factors \cite{sim80}.
%\begin{equation}
%G^M_{\tau}(q^2)=\mu_s^{\tau} [1+q^2/(18.23 \ {\rm fm}^{-2})]^{-2}\, .
%\end{equation}
In the convection currents, we use bare orbital $g$-factors, $g_{\ell}^p=1$ and 
$g_{\ell}^n=0$, scaled by $q$-dependent form factors given by a sum of monopoles
parametrized in Ref. \cite{sim80} for the proton and by the difference of two 
Gaussians \cite{chan76} for the neutron.
%\begin{equation}
%G^E_{p}(q^2)=\sum_{n=1}^4 \frac{a_n}{1+q^2/m_n^2}\, .
%\end{equation}
 
%\begin{equation}
%G^E_{n}(q^2)=\exp (-q^2r_{+}^2/4) - \exp (-q^2r_{-}^2/4)\, ,
%\end{equation}
%with $r_\mp ^2=0.507413\pm  0.038664 \ {\rm fm}^2$ .

The effects of Coulomb distortion could be evaluated quantitatively in the
distorted-wave Born approximation (DWBA) with a phase-shift calculation \cite{yen54}.
Nevertheless,  neglecting the Coulomb distortion offers clear advantages for the
analysis and interpretation of experimental data on magnetic scattering and PWBA is
commonly used. In PWBA the connection between data and the underlying physical
quantities is more transparent and calculations are simplified. The most important
effect of Coulomb distortion is accounted for by using an effective momentum transfer.
The general procedure is to convert the experimental form factors into plane-wave
form factors that can be directly compared with PWBA calculations, as was done
in Ref.  \cite{donnelly}.

\subsection{Mean field and form factors in a deformed formalism}

The ground state of an axially symmetric deformed nucleus is characterized by its
angular momentum $I$, its projection along the symmetry axis $k$, and parity $\pi$.
Initial and final states are the same for elastic scattering ($Ik^{\pi}$).

The magnetic  multipole form factors  $F^{M\lambda}$ can be written in terms of intrinsic 
form factors ${\cal F}^{M \lambda}$ weighted by angular momentum dependent coefficients.
To lowest order in an expansion in powers of the total angular momentum, the intrinsic 
multipoles depend only on the intrinsic structure of the ground-state band \cite{elvira}. 
The transition multipoles in Eq. (\ref{ftfm}) for the elastic case $I_f = I_i = k$ can 
be written as

\begin{eqnarray}
\left. F^{M\lambda} \right| _{\rm def}  
&=&\langle kk \lambda 0\ |kk\rangle {\cal F}^{M\lambda}_k 
+ \langle k\ \text{--}k\ \lambda \ 2k|kk\rangle {\cal F}^{M\lambda}_{2k} \nonumber \\
&&+\sqrt{\frac{\lambda(\lambda+1)}{2}} 
\langle kk\lambda 0 |kk\rangle {\cal F}^{M\lambda}_R \, .
\label{f_rot}
\end{eqnarray}
As we can see from the expression above, the magnetic form factors in odd-$A$ nuclei
receive two types of contributions, single-particle
(${\cal F}^{M\lambda}_{k}, {\cal F}^{M\lambda}_{2k}$) and collective (${\cal F}^{M \lambda}_R$).
The single-particle multipoles depend only on the single-particle intrinsic wave 
function of the odd nucleon if the even-even core is time-reversal invariant, as we 
assume in this work. They are different from zero only for $k\ne 0$ bands and are 
given by

\begin{eqnarray}
 {\cal F}^{M\lambda}_{k} & = &  \langle \phi_k | \hat{T} ^{M\lambda}_0 |
\phi_k \rangle \, , \label{fmk}\\
{\cal F}^{M\lambda}_{2k} & = &  \langle \phi_k | \hat{T} ^{M\lambda}_{2k} |
\phi_{\bar k} \rangle +\delta_{k,1/2} \frac{a}{\sqrt{2}} {\cal F}^{M\lambda}_{R} \, . 
\label{fm2k}
\end{eqnarray}
$\hat{T} ^{M \lambda}_{\mu}$ is the $\mu$ component of the $M \lambda$ tensor operator
(see Eq. (\ref{tensor}) and Ref. \cite{forest}).  $\phi_k$ and $\phi_{\bar k}$ are the 
intrinsic wave functions of the odd nucleon and its time reversed, respectively, and 
$a=<\phi_k|j_+|\phi_{\bar k}>$ is the decoupling parameter for $k=1/2$ bands.

${\cal F}^{M \lambda}_R$ are the magnetic multipoles of the collective rotational current
(rotational multipoles) that depend on the nuclear rotational model used to describe
the band \cite{elvira}. These contributions were studied \cite{ps,berdi} by using
different microscopic and macroscopic models, concluding that they are, in general,
small compared to single-particle contributions. They are only expected to be relevant
in the $M1$ multipoles at low $q$ and do not differ much from one rotational model to
another. Thus, we opted here for the cranking model, that produces better moments of
inertia.

The nuclear structure is described within a self-consistent axially symmetric deformed
HF+BCS formalism. The wave function for the $i$ single-particle state is written in
terms of the spin components $\phi_i^+$ and $\phi_i^-$ as \cite{vautherin},

\begin{eqnarray}
\phi_i({\bf R},\sigma) &=&  \phi_i^+ (r,z)
\exp(\mathrm{i}\Lambda^-\varphi) \chi_+(\sigma) \nonumber \\
&& + \phi_i^- (r,z) \exp(\mathrm{i}\Lambda^+\varphi) \chi_-(\sigma) .
\end{eqnarray}

The variables $r$, $z$ and $\varphi$ are the cylindrical coordinates of the 
radius-vector ${\bf R}$. $\chi_\pm(\sigma)$ are the spin wave functions and 
$\Lambda^\pm = \Omega_i \pm 1/2 \ge 0$, where $\Omega_i$ is the projection along the 
symmetry axis of the total angular momentum, and it characterizes the single-particle 
Hartree-Fock solutions for axially symmetric deformed nuclei, together with parity 
$\pi_i$. 

The wave functions $\phi_i$ are expanded into eigenfunctions
$\psi_\alpha ({\bf R},\sigma)$ of an axially deformed harmonic oscillator potential
using eleven major shells,

\begin{equation}
\phi_i({\bf R},\sigma)= \sum_\alpha C_\alpha^i \psi_\alpha ({\bf R},\sigma) \, ,
\end{equation}

with $\alpha=\left\lbrace n_r, n_z,\Lambda,\Sigma\right\rbrace $. All the results
presented in this work correspond to the Skyrme interaction SLy4 \cite{chabanat},
which has been thoroughly tested on many nuclear properties along the full nuclear
chart.

In the present work, the mean field of the odd-$A$ nucleus is generated within the 
equal filling approximation (EFA), a prescription used in self-consistent mean-field 
calculations for odd-$A$ nuclei that preserves time-reversal invariance. In this
approximation half of the unpaired nucleon sits in a given orbital and the other
half in the time-reversed partner. The odd nucleon orbital, characterized by
$\Omega_i=k$ and $\pi_i$, is chosen among those around the Fermi level, according to
the experimental ground-state spin and parity values. 
It is worth mentioning that with this choice of spin and parity  we obtain the minimum
of the energy in most cases. In the few cases where the minimum corresponds to
a different spin and parity, these assignments are found experimentally very close to
the ground states.

The explicit expressions for all the intrinsic form factors in Eqs. (\ref{fmk}) and  
(\ref{fm2k}) in terms of these wave functions can be found in \cite{elvira,kova}.
Expressions for the intrinsic rotational multipoles ${\cal F}^{M\lambda}_R$ can be
also found in  Ref. \cite{elvira} for different microscopic and macroscopic models.

Nuclei with quadrupole deformation parameters $| \beta_p | \leq 0.06$ are considered
in this work as spherical and the  spherical limit  of the present deformed formalism
is applied to them. In Tables \ref{table_isotopes} and \ref{table_isotones} these
cases are identified with the label `sph' in the column for quadrupole deformations.  
Thus, for these nuclei we first constrain the deformed calculation to zero deformation. 
In the spherical limit there are no collective magnetic multipoles
(${\cal F}^{M\lambda}_R=0$) and the form factors are related to the deformed ones by

\begin{equation}
\left. F^{M\lambda} \right| _{\rm sph\ limit} =
[\eta _j^{\lambda}]^{-1} \left. F^{M\lambda} \right| _{\rm def} \ ,
\label{sphlimit}
\end{equation}
with  $\left. F^{M\lambda} \right| _{\rm def}$  as in Eq. (\ref{f_rot}) and
$\eta _j^{\lambda}$ given by \cite{merino}

\begin{equation}
\eta _j^{\lambda} = \langle jj\lambda 0|jj  \rangle ^2
\left[ 1+\delta_{\lambda,2j}
\frac{ \langle j\ \text{--} j\ \lambda \ 2j|jj \rangle ^2}
{ \langle  jj \lambda 0 |jj\rangle ^2} \right] \ .
\label{eta_jl}
\end{equation}

It was found in \cite{merino} that the results from  a standard spherical formalism
agree perfectly with the results obtained in the spherical limit of the present
deformed formalism.

\subsection{Correlation effects in the spherical limit}

Many experimental nuclear data show sizable effects of nucleon-nucleon (NN) correlations
that cannot be accounted for within mean-field approximations  \cite{AHP,Kadrev96}.
However, the single-particle picture can be recovered with methods involving NN
correlations by using the natural orbital representation \cite{Lowdin55} of the
one-body density matrix (OBDM). The natural orbitals and occupation numbers are obtained
by diagonalizing the OBDM solving the equation:

\begin{equation}
\int d{\bf r^{\prime}}\rho({\bf r},{\bf r^{\prime}})\psi_{\alpha}({\bf r^{\prime}})=
n_{\alpha}\psi_{\alpha}({\bf r}),
\label{eq:diagonal}
\end{equation}
where $\psi_{\alpha}({\bf r})$ are the natural orbitals and $n_{\alpha}$ are the natural
occupation numbers.

In the present work, we include correlations within the Coherent Density Fluctuation
Model (CDFM) \cite{Antonov79,AHP,Antonov94,Kadrev96}. The OBDM in this model has the
form:
\begin{equation}
\rho({\bf r},{\bf r^{\prime}})=\int |{\cal F}(x)|^{2}
\rho_{x}({\bf r},{\bf r^{\prime}})dx,
\label{eq:cdfm}
\end{equation}
where

\begin{equation}
\rho_{x}({\bf r},{\bf r^{\prime}})=3\rho_{0}(x) \frac{j_{1}(k_{F}(x)|{\bf r}-
{\bf r^{\prime}}|)}{k_{F}(x)|{\bf r}-{\bf r^{\prime}}|} 
 \Theta \left (x-\frac{|{\bf r}+{\bf r^{\prime}}|}{2}\right ) \, ,
\label{eq:flucton}
\end{equation}
with

\begin{equation}
\rho_{0}(x)=\frac{3A}{4\pi x^{3}}  \, , \quad
k_{F}(x)=\left(\frac{3\pi^{2}}{2}\rho_{0}(x)\right )^{1/3}\, .
\label{eq:fermi}
\end{equation}
The weight function $|{\cal F}(x)|^{2}$ in (\ref{eq:cdfm}) in the case of 
monotonically-decreasing density distributions can be obtained from the density as
\begin{equation}
|{\cal F}(x)|^{2}=-\frac{1}{\rho_{0}(x)} \left. \frac{d\rho(r)}{dr}\right |_{r=x} .
\label{eq:weight}
\end{equation}

The effect of these correlations has been studied in previous works
\cite{merino,Kadrev96}, showing that they shift the tails of the form
factors to higher momentum transfer, improving the agreement with experiment.
We have included these correlations in the calculations on the magnetic form
factors in spherical nuclei. Although not shown here, we have also
calculated the correlations on the spherical limit of the deformed nuclei, finding
a similar effect in the tails of the magnetic form factors.
As we shall see, this effect does not change the clear differences observed in the
profiles of the form factors from different  $I^{\pi}$ and deformations that manifest
themselves mainly at low momentum transfer.

\section{Results and discussion }

In this section we study several isotopic and isotonic chains that have been chosen
according to the criteria discussed above. We first consider vanadium and aluminum
isotopes as examples of the general expected behavior within isotopic chains with
a fixed number of odd protons and a variable number of even neutrons. In principle,
one expects to have the same spin and parity for all the members of the chain, as well
as a similar deformation, which in our case is spherical for vanadium and prolate for
aluminum isotopes. As we shall see, the profiles of the calculated magnetic form
factors are quite similar as well. On the other hand, the isotopic chains of fluorine,
manganese, iodine, and cesium under study here are characterized by a change in the
angular momentum of the ground states related to a shape transition. The corresponding
magnetic form factors also exhibit a great sensitivity to these structural changes.
The isotonic chains with $N=9,11,25,$ and 57 are also chosen as examples with angular
momentum changes and the magnetic form factors show also very different profiles.

\subsection{Static moments}

Tables \ref{table_isotopes} and \ref{table_isotones} show theoretical and experimental
ground-state properties of the considered isotopic and isotonic chains, respectively.
Namely, spins and parities, half-lives, charge root-mean-square radii $r_c$, quadrupole
proton deformations $\beta_p$, spectroscopic  electric quadrupole moments $Q_{\rm lab}$,
and magnetic moments $\mu$. The results are compared with experimental data from
\cite{angeli} for radii and from \cite{stone} for quadrupole and magnetic moments.
Note that the lack of a sign in the experimental $Q$ and $\mu$ means that it is
still undetermined.

The relationship between the intrinsic quadrupole moment $Q_0$ and the quadrupole
deformation parameter $\beta_p$ is given by
\begin{equation}
Q_0 = \sqrt{\frac{5}{\pi}} Ze \langle r_p^2 \rangle \beta_p \, ,
\end{equation}
where $<r_p^2>$ is the proton mean-square radius. The measured quadrupole moment 
$Q_{\rm lab}$ is related to the intrinsic quadrupole moment $Q_0$ by
\begin{equation}
Q_{\rm lab} = \frac{3k^2 - I(I+1)}{(I+1)(2I+3)} e Q_0 \, .
\end{equation}
The measured electric quadrupole moments in the tables correspond to ground states
$I=k$. Note that in the cases $I=k=1/2$, $Q_{\rm lab} = 0$.

The magnetic moments in the deformed case $\mu_{\rm th, def}$ are obtained from the 
expression,

\begin{eqnarray}
\mu_I &=& g_R I + \frac{k^2}{I+1} \big[ g_k - g_R +  \nonumber \\
  && \delta_{k,1/2} (2I+1)(-1)^{I+1/2} \sqrt{2} g_{2k} \big] \, ,
\label{mui}
\end{eqnarray}
where $g_R$ is approximated by $Z/A$ and  $g_k$, $g_{2k}$ are defined in Ref.
\cite{elvira}. We also show for comparison the Schmidt values $\mu_{\rm th, sph}$ 
obtained in the spherical limit.

Similarly to the cases studied in \cite{merino}, we find here that the discrepancies
between the calculated and the measured $r_c$ are just of a few per thousand in most
cases. Only in the lightest nuclei the errors are somewhat larger, but always below
two percent. In the case of the isotopic chains we can see that the charge radii
increase slightly as the number of neutrons increase. This is a general trend expected
from the tendency of protons to overlap maximally with neutrons and thus spreading the
spatial distribution of the latter (see e.g., Ref. \cite{antonov_ff}). This general
trend is nevertheless altered by local effects related to deformation. A clear example
of this peculiar behavior can be seen in the manganese isotopes, where the shape changes
from prolate in $^{51}$Mn to spherical in the magic ($N=28$) $^{53}$Mn, and again to
prolate in $^{55}$Mn. This structural change makes the charge radii decrease from
$^{51}$Mn to  $^{53}$Mn and increase from  $^{53}$Mn to  $^{55}$Mn. 
In the case of the isotonic chains, as compared with the isotopic chains, we observe
a more dramatic increase of the charge radii as the number of protons increases. This
is expected as the protons occupy outer orbitals.

The quadrupole moments are calculated for the deformed nuclei and compared with the  
measured $Q_{\rm lab}$. They agree in sign and magnitude.

Concerning magnetic moments,  we show the Schmidt values as well as the values from
Eq. (\ref{mui}) in the deformed cases. We see that the Schmidt values reproduce
reasonably well the magnetic moments of spherical nuclei, whereas for the deformed
nuclei there is a clear improvement of the agreement with experiment using the
deformed formalism.

\subsection{Form factors in isotopic chains}

In the case of even-$Z$ and odd-$N$ isotopes the spin and parity of the ground
state, which depend on the odd-neutron state, change from isotope to isotope. 
Therefore, we expect rather different form factors as we move in an isotopic chain
of this kind. On the contrary, in the case of even-$N$ and odd-$Z$ isotopes that we
study in this work, the spin/parity of the ground state remain unchanged as long as
there are no critical changes in the nuclear structure. The magnetic form factors of
these isotopes are expected to be similar because they are mainly determined by the
wave function of the odd proton, that does not change much. Therefore, changes in the
spin/parity of the isotopes in these chains become signatures of structural transitions
induced by the collective effect of the different number of neutrons on the odd-proton
wave function. As it will be shown here, these effects can also be studied through
the magnetic form factors.

The spherical limit defined earlier is applied to those nuclei with very small
$\beta_p$ values. This limit involves a spherical constraint of the mean-field
calculation, a redefinition of the multipole form factors given by Eq. (\ref{sphlimit}),
and the removal of collective contributions ${\cal F}^{M\lambda}_R$ from rotations of
the core. 

In the following figures for the isotopic chains under study in this work, we show in
the top panels (a) a comparison of the total magnetic form factors corresponding to the
three nuclei considered in each chain. In the lower panels (b), (c), and (d) we show,
for each member of the chain, the total magnetic form factor decomposed into the
contributing multipolarities. In the case of spherical nuclei we include the results
of the CDFM calculations in the total magnetic form factor.

\vspace{0.5cm}
{\centerline {\bf Vanadium isotopes}}
\vspace{0.5cm}

\begin{figure*}[h]
%\begin{figure}[H]
\centering
\includegraphics[width=140mm]{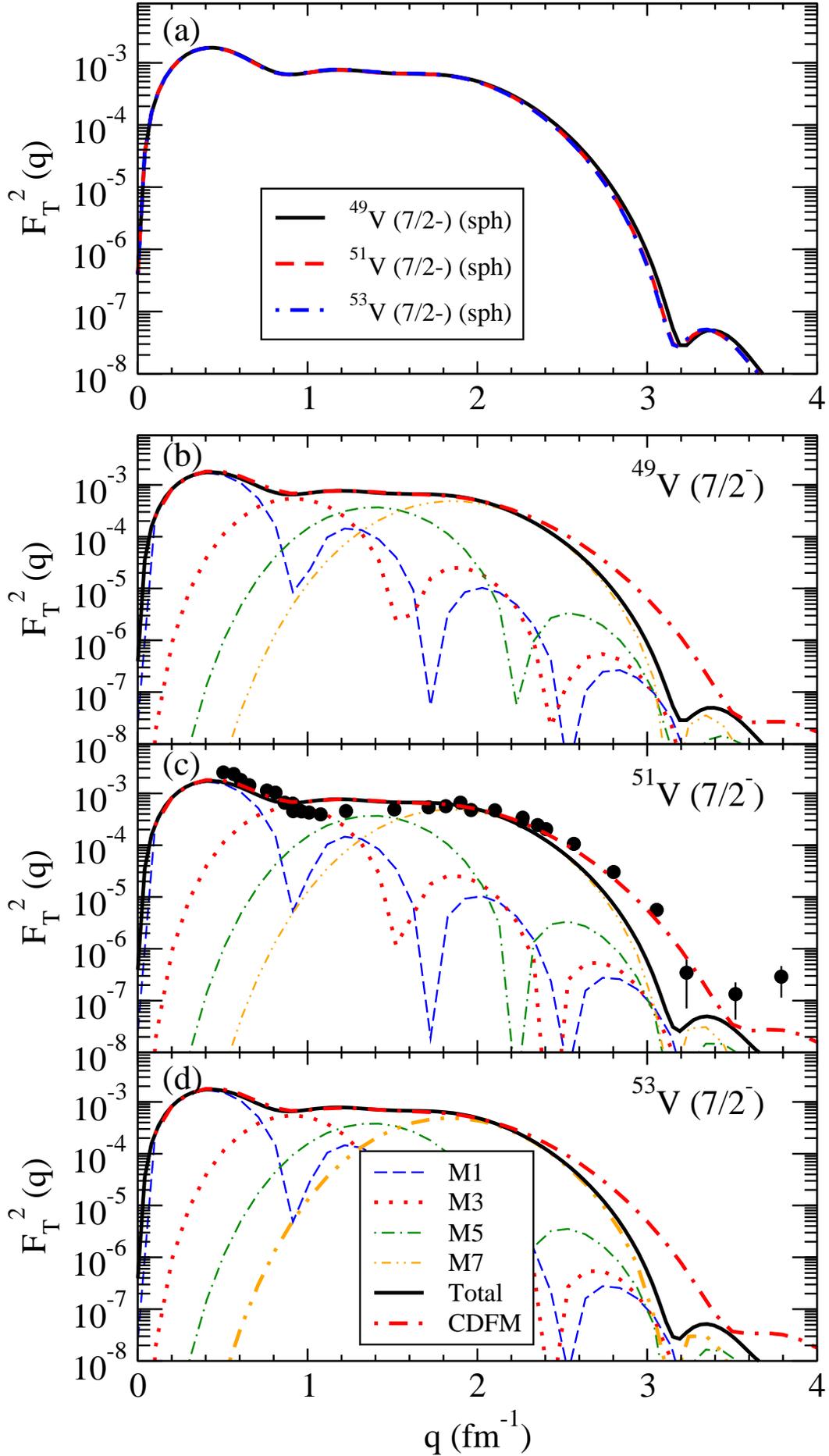}
\caption{
  Total magnetic form factors for the $Z=23$ isotopes $^{49,51,53}$V (a) and their decomposition
  into the contributing multipoles for $^{49}$V (b), $^{51}$V (c), and $^{53}$V (d).
  CDFM calculations are shown for the spherical cases. Data are taken from
  \cite{donnelly,arita81,platchkov83}}
\label{fig_v}
\end{figure*}

In Fig. \ref{fig_v}  we show the magnetic form factors in isotopes of vanadium ($Z=23$),
$^{49,51,53}$V. They are examples of three spherical isotopes with the same observed
ground-state spins and parities (see Table I). They are $I^{\pi}=7/2^-$ odd-$A$ and
odd-$Z$ nuclei, where the valence proton sits in the $f_{7/2}$ shell. Because of the
similar nuclear structure in the three isotopes, the magnetic form factors in
Fig. \ref{fig_v}  are quite similar and therefore this chain is an example
of the expected behavior of odd-$Z$ even-$N$ isotopes. Panels (b), (c), and (d) show
the multipole decomposition of the total magnetic form factors, where the lowest
multipole ($M1$) determines the low-$q$ behavior of the magnetic form factors, whereas
the highest multipole, $M7$ in this case, determines the large-$q$ tails. The other
multipoles contribute in different degrees to the intermediate region of momentum
transfer. This is generally true for all the cases under study. The experimental data
on  $^{51}$V from \cite{donnelly,arita81,platchkov83} are well reproduced by the
calculations.

Nucleon-nucleon correlations included in the CDFM approach are shown in the three
spherical isotopes. They influence the form factor beyond $q=2$ fm$^{-1}$ and the
main effect is to increase the form factor at high momentum transfer. The agreement
with experiment in the case of $^{51}$V is clearly improved by these correlations.

\vspace{0.5cm}
{\centerline {\bf Aluminum isotopes}}
\vspace{0.5cm}

\begin{figure*}[h]
%\begin{figure}[H]
\centering
\includegraphics[width=140mm]{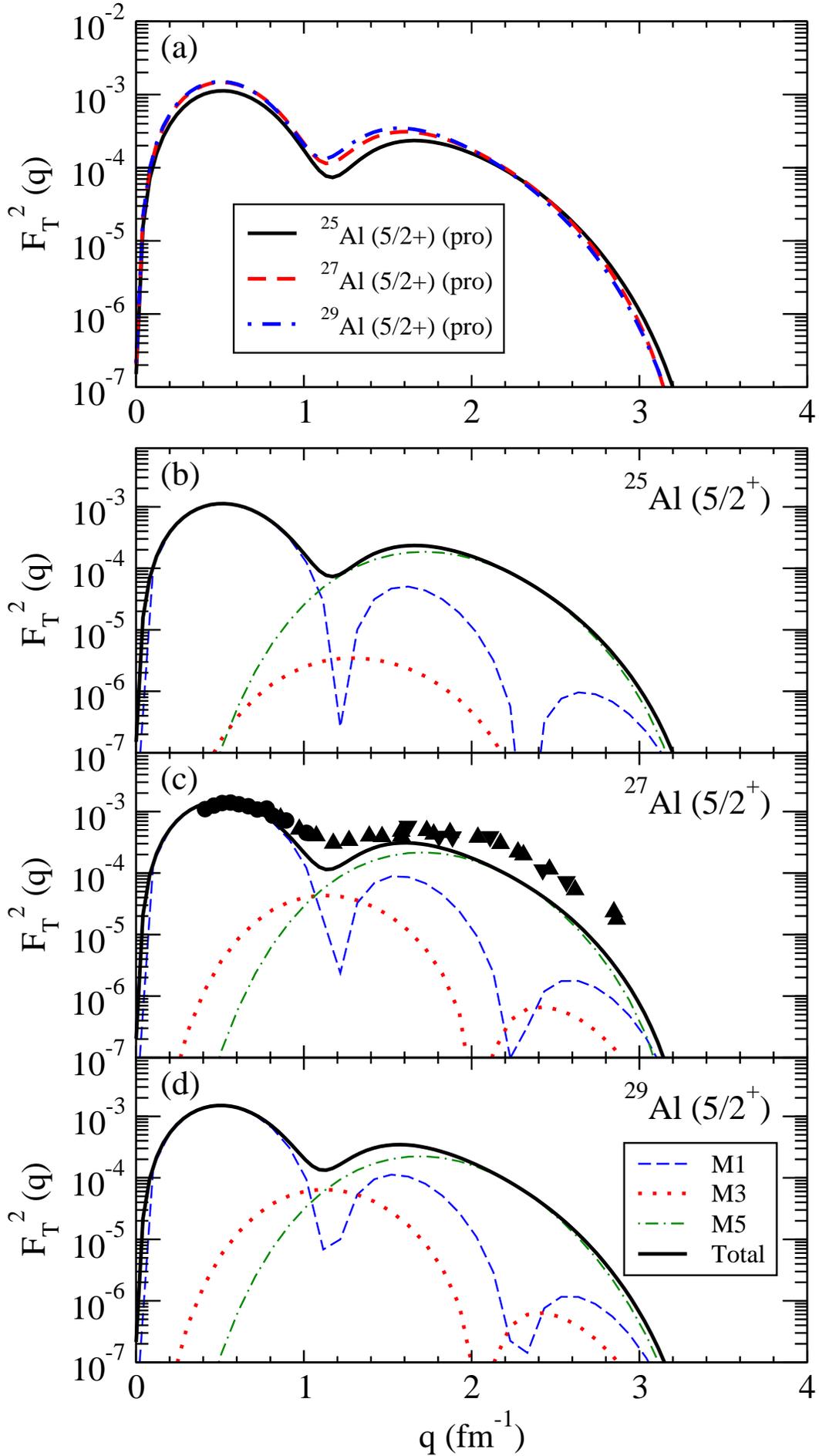}
\caption{Same as in Fig. \ref{fig_v}, but for the $Z=13$ isotopes $^{25,27,29}$Al.
  Data are taken from \cite{donnelly,lapikas73,li}
}
\label{fig_al}
\end{figure*}

In Fig. \ref{fig_al}  we show the magnetic form factors in isotopes of aluminum 
$(Z=13)$, $^{25,27,29}$Al. The three isotopes are found experimentally to have
$I^{\pi}=5/2^+$ in their ground states with the odd proton occupying the $d_{5/2}$
shell in the spherical limit. In the deformed model the odd proton occupies a
$k^{\pi}=5/2^+$ prolate state or a $k^{\pi}=1/2^+$ oblate state. Because the former
is energetically favored in the three isotopes, our description corresponds to a
prolate shape with the odd proton in a Nilsson-like state with asymptotic quantum
numbers $[202]5/2$. The three isotopes have a similar deformation (see Table I)
and the magnetic form factors are practically indistinguishable. Thus, adding
couples of neutrons in this isotopic chain does not change the structure of the
odd-proton state. This chain is another example of the expected behavior of odd-$Z$
even-$N$ isotopes, but in this case for deformed nuclei. The multipole decomposition
shows that the multipole $M1$ determines the behavior below $q$=1 fm$^{-1}$ and the
multipole $M5$ determines the behavior beyond, whereas the intermediate multipole $M3$
is irrelevant.

Experimental data for $^{27}$Al from   \cite{donnelly,lapikas73,li} are well reproduced,
except in the high-$q$ region where the data are underestimated due to the lack of
short-range correlations.

\vspace{0.5cm}
{\centerline {\bf Fluorine isotopes}}
\vspace{0.5cm}

The case of fluorine isotopes ($Z=9$), $^{17,19,21}$F, is different and much more 
interesting from a nuclear structure point of view. Experimentally, the ground states
of  $^{17}$F and $^{21}$F have $I^{\pi}=5/2^+$, but  $^{19}$F has $I^{\pi}=1/2^+$. In a
simple Nilsson diagram one can see that the odd proton sits on the $d_{5/2}$
spherical shell either on the orbital $k^{\pi}=5/2^+$ in the oblate region or on the
orbital $k^{\pi}=1/2^+$  in the prolate region. Since a prolate configuration is favored
energetically in $^{19}$F, the odd proton is expected to occupy the Nilsson orbital
$[220]1/2$. On the other hand, the oblate shape is preferred in $^{21}$F, and the odd
proton occupies in this case the Nilsson orbital $[202]5/2$.

\begin{figure*}[h]
%\begin{figure}[H]
    \centering
\includegraphics[width=140mm]{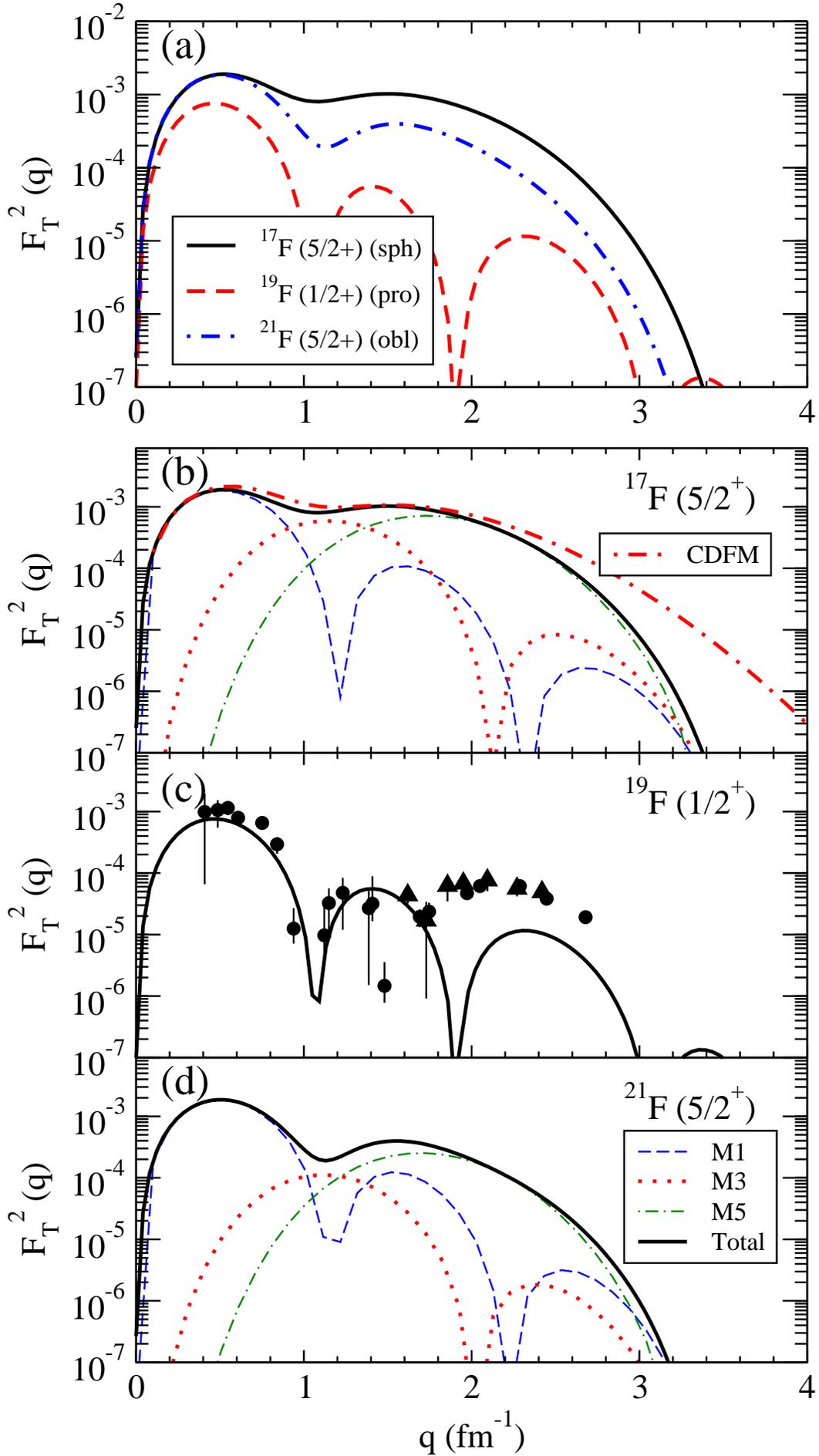}
\caption{Same as in Fig. \ref{fig_v}, but for the $Z=9$ isotopes $^{17,19,21}$F.
 Data are taken from \cite{williamson,donne86}
}
\label{fig_f}
\end{figure*}

In the top panel of Fig. \ref{fig_f} we can see the total magnetic form factors of
the three isotopes together. A clear difference can be observed among them related
to the effect of deformation between the spherical $^{17}$F and the deformed $^{19,21}$F,
but also related to the different angular momentum of the ground states that gives rise
to different multipole contributions. In  panels (b), (c), and (d) we can see
separately the multipole contributions in the cases of $^{17}$F, $^{19}$F and $^{21}$F,
respectively. The data in $^{19}$F are from  Refs. \cite{williamson,donne86}. The only
multipole is $M1$ and exhibits a three-peaked structure in this region of momentum
transfer, which seems to be also the case with the data.

The different $I^{\pi}$ of the ground states observed as we move in the isotopic chain
indicates a structural change. In the deformed picture this effect is associated with
a shape transition from spherical ($5/2^+$) in  $^{17}$F to prolate ($1/2^+$) in
$^{19}$F and to oblate  ($5/2^+$) in  $^{21}$F. Thus, the neutron environment of the
three isotopes, $N=8,10,12$, determines the state where the odd proton sits. The
self-consistent deformations in these isotopes can be seen in Table I. This structural
change is translated into the magnetic form factor, which is indeed sensitive to the
odd-proton state. The result is that in $^{17}$F and $^{21}$F, the multipoles $M3$ and
specially $M5$ contribute by filling the high-$q$ region and a two-peaked structure
is obtained, rather different to the  profile of the $^{19}$F isotope that exhibits a
three-peaked structure.

The configurations with  $1/2^+$ and $5/2^+$ in the three isotopes are indeed quite
close in energy, competing to each other for being ground states. It is worth noting
that experimentally, a $5/2^+$ excited state is observed in $^{19}$F at an energy
$E=197$ keV. Similarly, $1/2^+$ excited states are observed in both $^{17}$F and
$^{21}$F at energies $E=495$ keV and $E=280$ keV, respectively. These features point at
a shape coexistence between spherical, oblate, and prolate configurations or to a
mixture of them. Given the sensitivity of the magnetic form factors to the details of
the odd-proton wave function, elastic scattering experiments could be used to gain
information on these properties.

Pure single-particle results produce already a good agreement with experiment in the
first peak of $^{19}$F. Core contributions may not be very reliable in such a light
nucleus, but we have included them in the rigid rotor model that improves slightly
the agreement with experiment. Our calculations reproduce the structure of three
peaks of the form factor measured although the third peak is underestimated. However,
the nucleon-nucleon correlations calculated in the spherical nucleus  $^{17}$F that
increase the form factors beyond $q=2$ fm$^{-1}$ suggest that a similar effect would
be expected in the deformed nucleus $^{19}$F improving the agreement with experiment
in the region of the third peak.

\vspace{0.5cm}
{\centerline {\bf Manganese isotopes}}
\vspace{0.5cm}

The case of manganese isotopes ($Z=25$), $^{51,53,55}$Mn, is also an interesting example
of a shape transition, which is already anticipated by the change in the angular
momentum of their ground states, namely, $5/2^-$ in $^{51}$Mn, $7/2^-$ in $^{53}$Mn, and
$5/2^-$ in $^{55}$Mn. This change is related to a shape transition from prolate in
$^{51}$Mn to spherical in $^{53}$Mn (associated to the magic number $N=28$) and again to
prolate in $^{55}$Mn, as can be seen in Table I. The odd proton occupies the spherical
shell $f_{7/2}$, while the prolate configurations in  $^{51}$Mn and  $^{55}$Mn correspond
to a Nilsson-like state with asymptotic quantum numbers $[312]5/2$.

Low-lying $7/2^-$ excited states are observed experimentally in $^{51}$Mn and $^{55}$Mn
at $E=237$ keV and $E=126$ keV, respectively. A $5/2^-$ excited states is also found in
$^{53}$Mn at $E=378$ keV, showing again a competition between different configurations.

\begin{figure*}[h]
%\begin{figure}[H]
    \centering
\includegraphics[width=140mm]{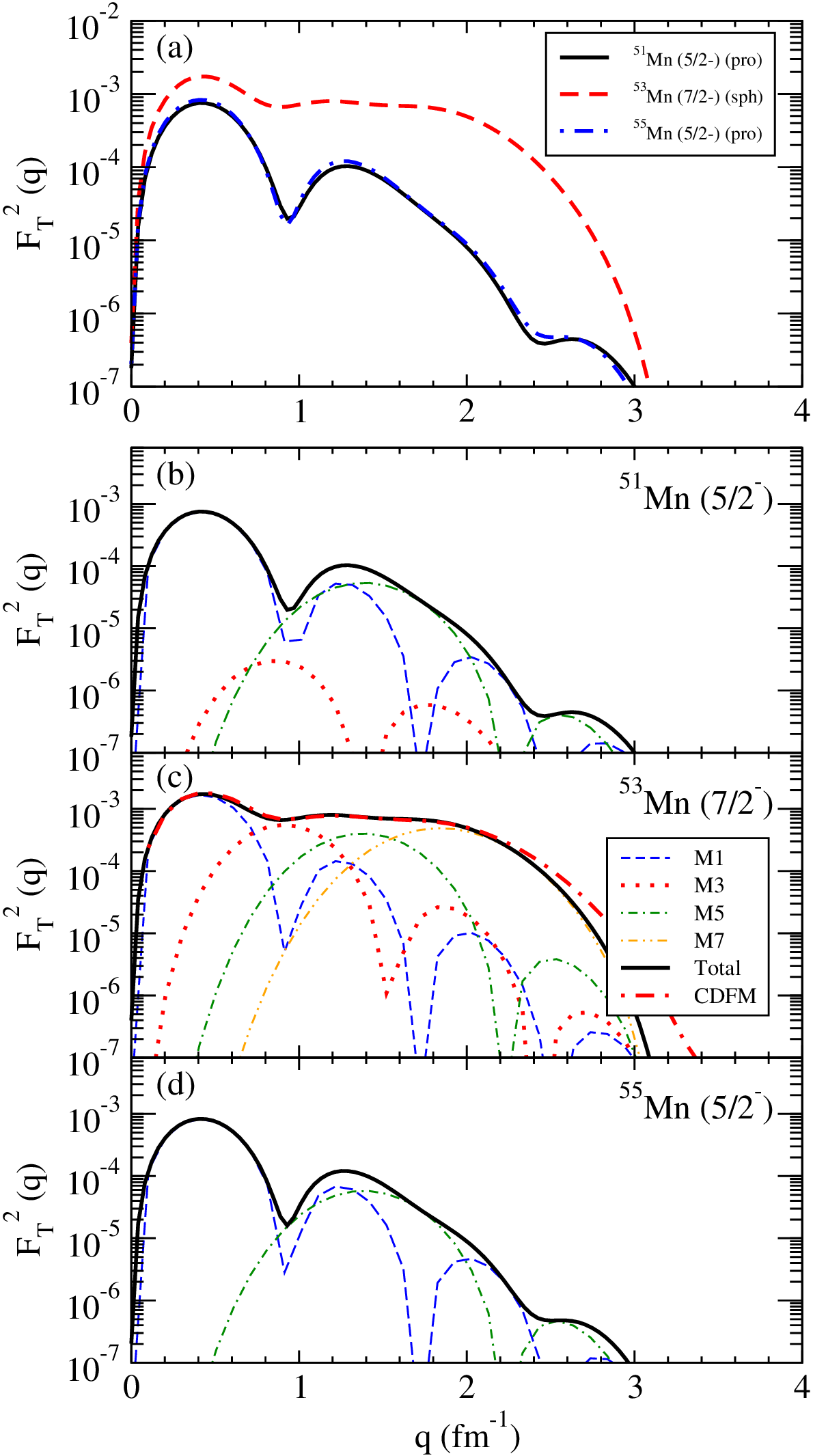}
\caption{Same as in Fig. \ref{fig_v},  but for the $Z=25$ isotopes $^{51,53,55}$Mn.
}
\label{fig_mn}
\end{figure*}

In the top panel (a) of Fig. \ref{fig_mn} we can see the total magnetic form factors
of the three isotopes, where a very clear difference between the spherical ($7/2^-$)
$^{53}$Mn isotope  and the prolate deformed  ($5/2^-$) $^{51,55}$Mn isotopes is observed.
Whereas the deformed isotopes $^{51,55}$Mn show a first peak at  $q=0.4$ fm$^{-1}$ and
a second peak, one order of magnitude smaller, centered at  $q=1.2$ fm$^{-1}$.
The spherical isotope $^{53}$Mn shows a small peak at $q=0.4$ fm$^{-1}$ followed by a
plateau-like structure up to about  $q=2$ fm$^{-1}$. Therefore, this is a very clear
example of a structural change to which the magnetic scattering will be extremely
sensitive.

The CDFM calculation on the spherical isotope  $^{53}$Mn shows again an increase of
the tails of the form factors at high momentum transfer.

\vspace{0.5cm}
{\centerline {\bf Iodine isotopes}}
\vspace{0.5cm}

\begin{figure*}[h]
%\begin{figure}[H]
    \centering
\includegraphics[width=140mm]{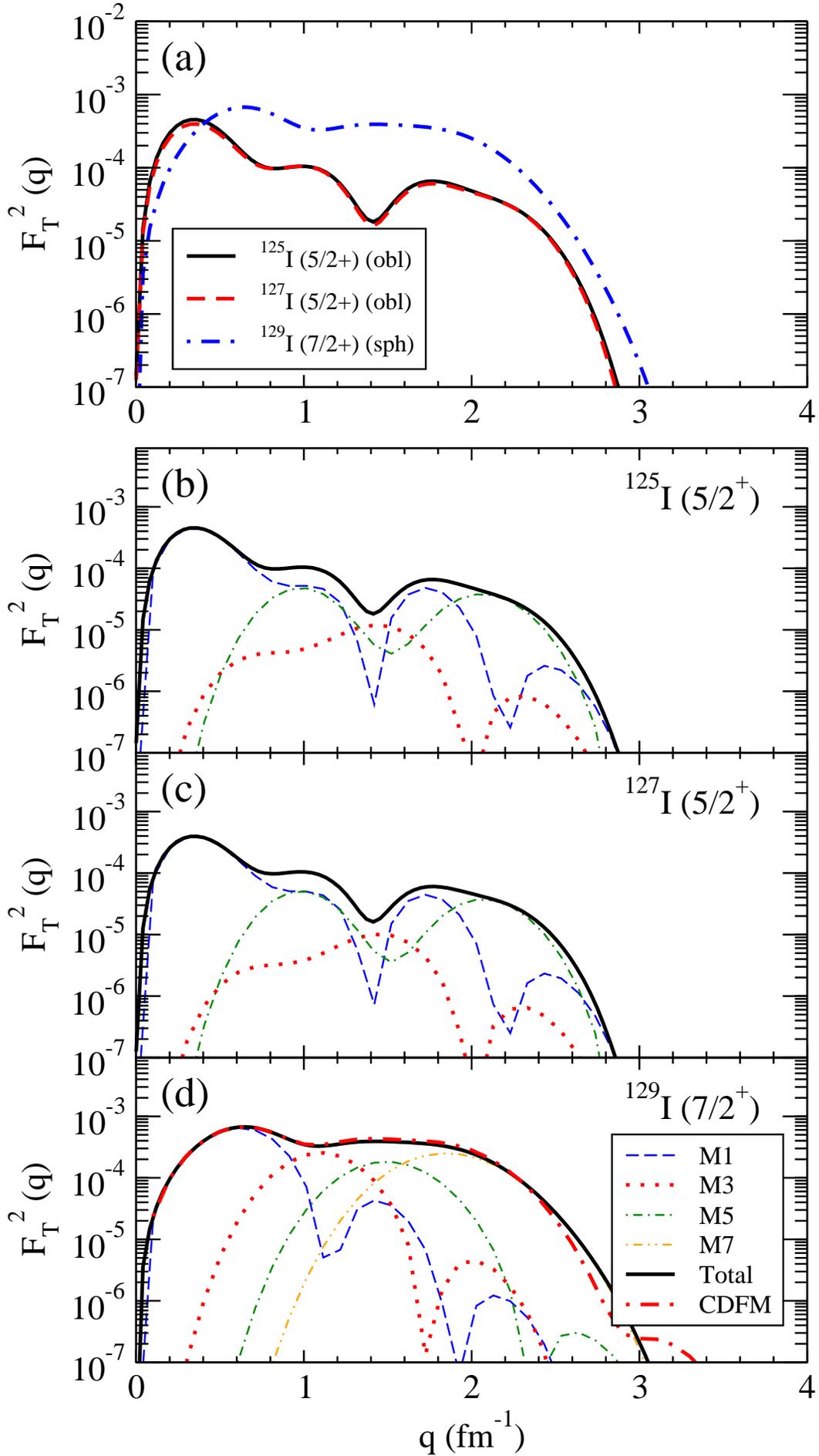}
\caption{Same as in Fig. \ref{fig_v},  but for the $Z=53$ isotopesr $^{125,127,129}$I.
}
\label{fig_i}
\end{figure*}

The magnetic form factors of iodine isotopes ($Z=53$) $^{125,127,129}$I are plotted in
Fig.  \ref{fig_i}. The measured angular momenta  are $5/2^+$ in $^{125}$I, $5/2^+$ in
$^{127}$I, and $7/2^+$ in $^{129}$I. This change is related to a shape transition from
oblate in $^{125,127}$I to spherical in $^{129}$I, as can be seen in Table I.
Excited states $7/2^+$ in $^{125}$I, $7/2^+$ in $^{127}$I, and $5/2^+$ in $^{129}$I are
also found at energies $E=114$ keV, $E=58$ keV, and $E=28$ keV, respectively.
The odd proton occupies the spherical shell $g_{7/2}$ in $^{129}$I, while it occupies
the Nilsson oblate state $[413]5/2$ in  $^{125,127}$I.

The magnetic form factor in $^{129}$I shows a smoother profile than the form factors of
$^{125,127}$I, which is caused by the enhancement of the intermediate multipoles in the
spherical case.
CDFM results are  shown in $^{129}$I, but they are very similar to the mean field results.

\vspace{0.5cm}
{\centerline {\bf Cesium isotopes}}
\vspace{0.5cm}

\begin{figure*}[h]
%\begin{figure}[H]
    \centering
\includegraphics[width=140mm]{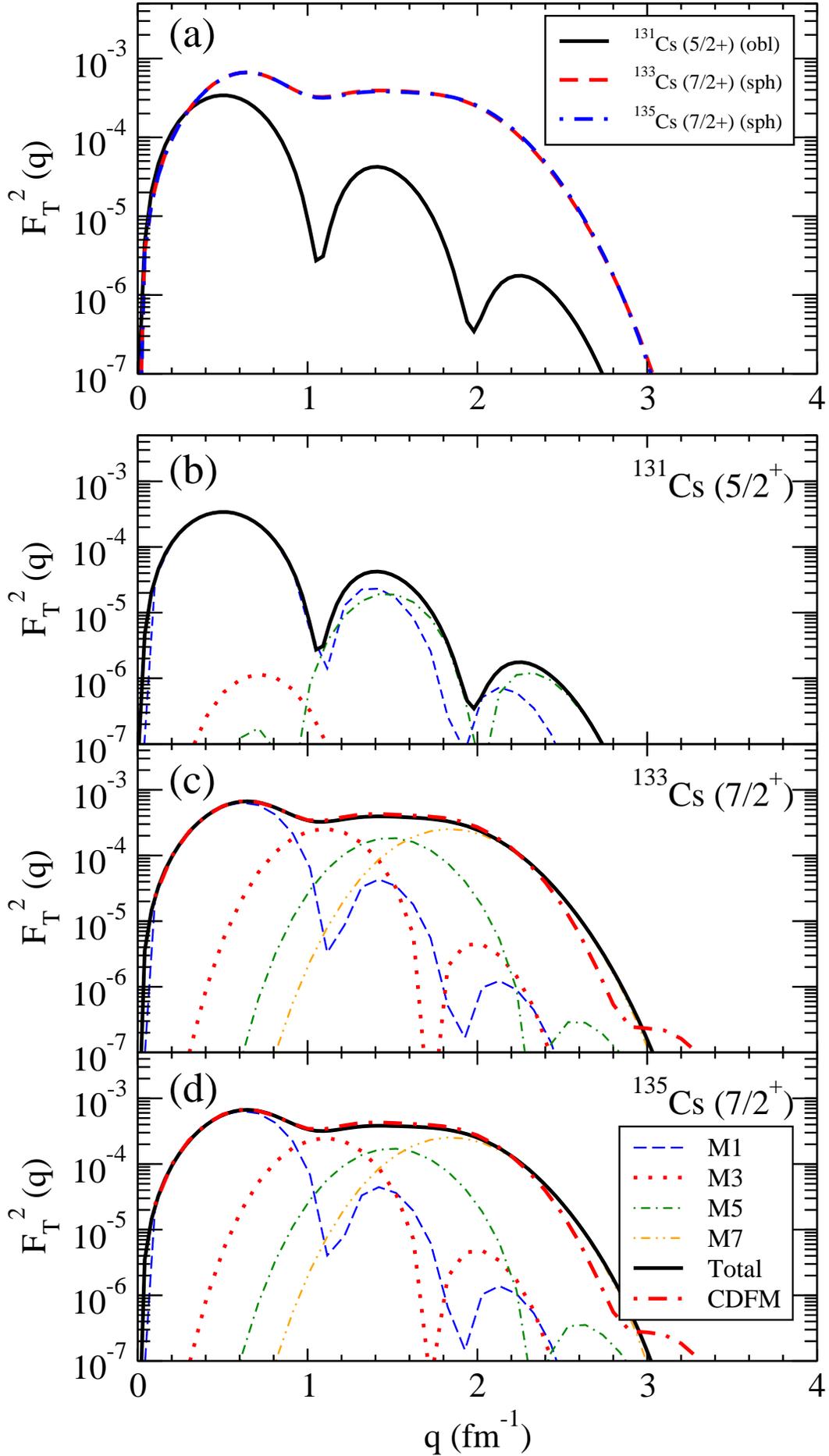}
\caption{Same as in Fig. \ref{fig_v},  but for the $Z=55$ isotopes $^{131,133,135}$Cs.
  }
\label{fig_cs}
\end{figure*}

The magnetic form factors of cesium isotopes ($Z=55$) $^{131,133,135}$Cs are plotted in
Fig.  \ref{fig_cs}. The experimental angular momentum and spin assignments of the
ground states are $5/2^+$ in $^{131}$Cs, $7/2^+$ in $^{133}$Cs, and $7/2^+$ in $^{135}$Cs.
This change is related to a shape transition from oblate  in $^{131}$Cs to spherical
in $^{133,135}$Cs, as can be seen in Table I. The spherical shell is now $g_{7/2}$
and the oblate Nilsson state is $[402]5/2$.

A low-lying excited state $7/2^+$ appears at $E=79$ keV in  $^{131}$Cs, while $5/2^+$
excited states appear at $E=81$ keV and $E=250$ keV in $^{133}$Cs and $^{135}$Cs,
respectively.

Similarly to the previous case, Fig.  \ref{fig_cs} shows a clear difference between the
magnetic form factors of the spherical and deformed cases. The form factor in $^{131}$Cs
shows a clear three-peaked structure, while the form factor in the spherical isotopes
$^{133,135}$Cs is rather smooth because of the enhancement of the intermediate multipoles.
As in the case of iodine, the CDFM calculations show little effect on cesium isotopes.

\subsection{Form factors in isotonic chains}

We consider now several odd-$N$ and even-$Z$ isotonic chains, which are characterized
by a change in the spin/parity of their ground states. Namely, we study the following
chains: i) $N=9$ with $^{15}$C, $^{17}$O, and $^{19}$Ne, ii) $N=11$ with $^{19}$O, $^{21}$Ne,
and $^{23}$Mg, iii) $N=25$ with $^{45}$Ca, $^{47}$Ti, and $^{49}$Cr, and iv) $N=57$ with
$^{99}$Mo, $^{101}$Ru, and $^{103}$Pd.

As we shall see, the differences in the magnetic form factors in a given chain are in
general larger between deformed isotones with the same $I^{\pi}$ than between deformed
isotopes with the same  $I^{\pi}$. This is due to the convection current changes
induced by the different number of protons in the members of an isotonic chain.

\vspace{0.5cm}
{\centerline { \textbf {\textit N = 9 isotones}}}
\vspace{0.5cm}

\begin{figure*}[h]
%\begin{figure}[H]
    \centering
\includegraphics[width=140mm]{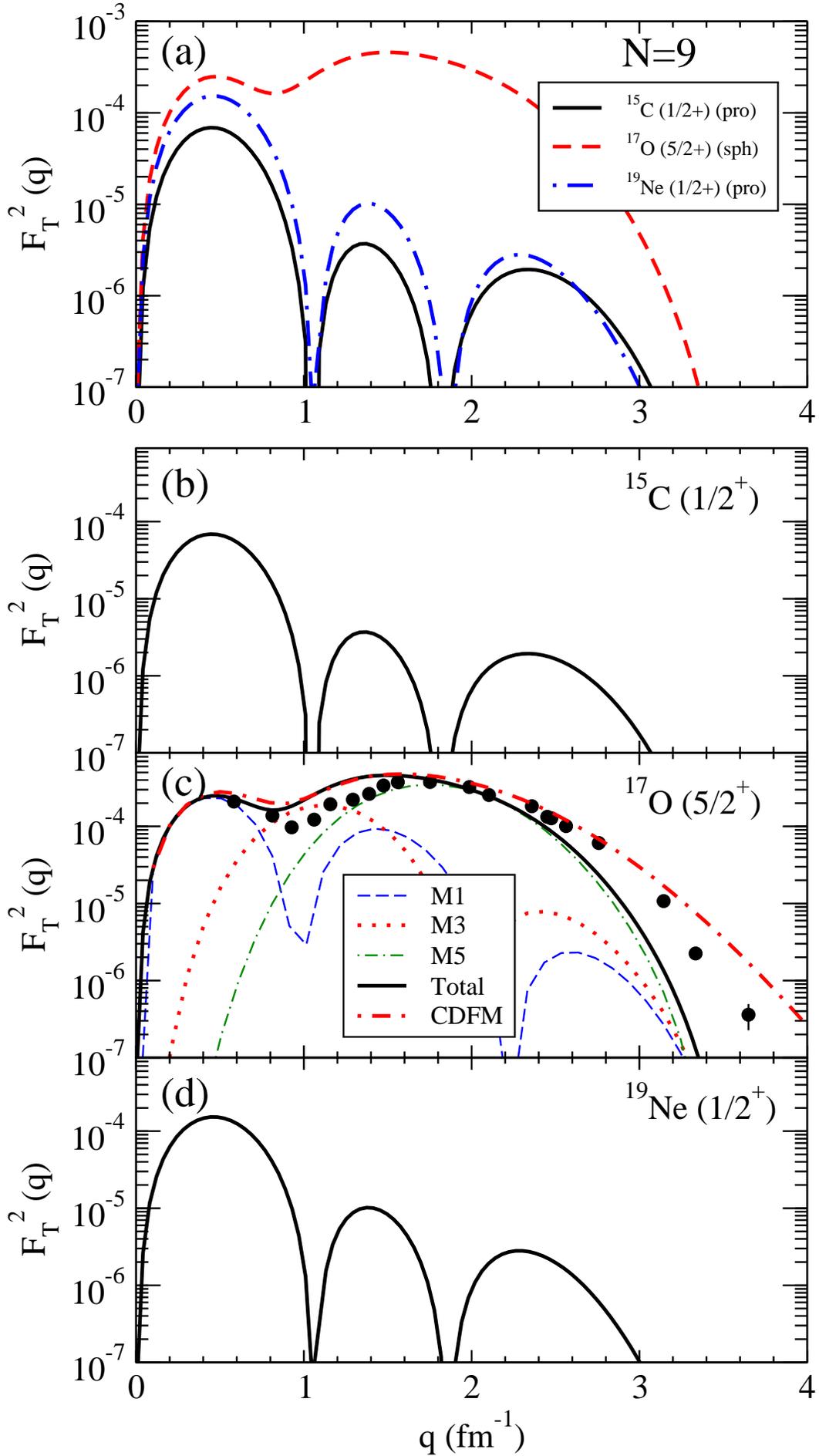}
\caption{Same as in Fig. \ref{fig_v},  but for the $N=9$ isotones $^{15}$C, $^{17}$O, and $^{19}$Ne.
  Data are taken from \cite{donnelly,hynes}.
}
\label{fig_n9}
\end{figure*}

We show in Fig. \ref{fig_n9} the chain of $N=9$ isotones given by $^{15}$C, $^{17}$O,
and $^{19}$Ne. The measured spin/parities are $1/2^+$ in $^{15}$C, $5/2^+$ in $^{17}$O,
and $1/2^+$ in $^{19}$Ne. This change is related to a shape transition from a prolate
deformation in $^{15}$C to a spherical shape in  $^{17}$O (related to the $Z=8$ magic
number), and to a prolate shape again in  $^{19}$Ne, as can be seen in Table I.

The odd neutron sits naturally in the $d_{5/2}$ shell in the spherical nucleus $^{17}$O,
but in the deformed isotones   $^{15}$C and  $^{19}$Ne the odd neutron sits in the
asymptotic orbital [220]1/2. It is also worth noting that $1/2^+$ excited states are
found experimentally in $^{17}$O at $E=871$ keV, and $5/2^+$ excited states are found
in $^{15}$C and  $^{19}$Ne at $E=740$ keV and  $E=238$ keV, respectively.

Fig. \ref{fig_n9} shows a  large  difference between the magnetic form factors
of these isotones. Whereas differences between the deformed nuclei $^{15}$C and
$^{19}$Ne, which are mainly due to the different contributions from the $Z=6$ and
$Z=10$ cores,  are not very significant, the difference with respect to the spherical
nucleus  $^{17}$O are dramatic and a clear signature of a shape transition. The
difference between the form factors for these  spherical or deformed isotones is more
than one order of magnitude beyond $q=1$ fm$^{-1}$.

Our results reproduce very reasonably the measured magnetic form factor in $^{17}$O
\cite{donnelly,hynes}, especially when correlations calculated within the CDFM
are included, that improve the agreement at high momentum transfer. It will be very
interesting to check against experiment the predictions for the form factors of
those isotones.

\vspace{0.5cm}
{\centerline { \textbf {\textit N = 11 isotones}}}
\vspace{0.5cm}

\begin{figure*}[h]
%\begin{figure}[H]
    \centering
\includegraphics[width=140mm]{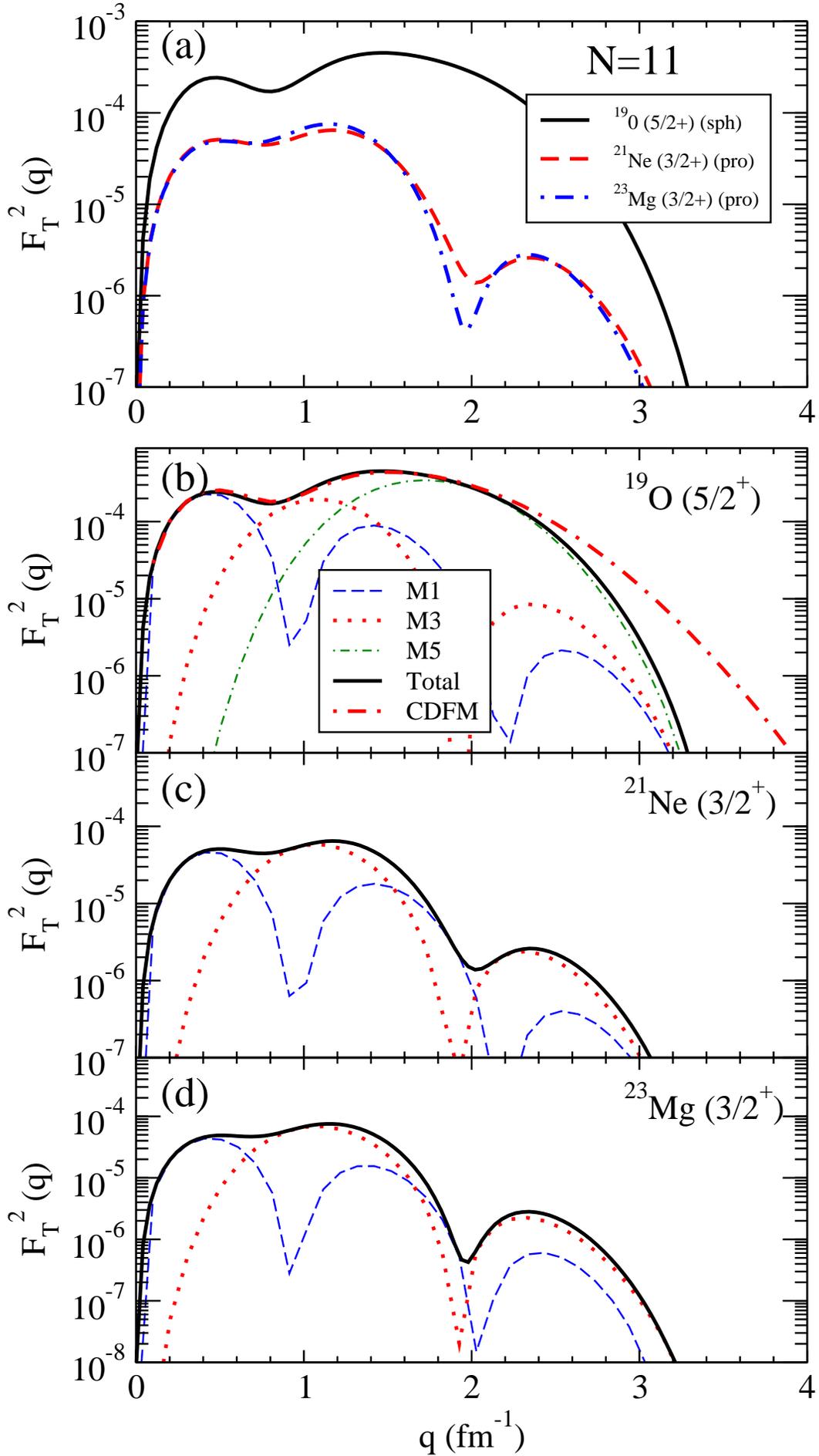}
\caption{Same as in Fig. \ref{fig_v},  but for the $N=11$ isotones  $^{19}$O, $^{21}$Ne, and $^{23}$Mg.
}
\label{fig_n11}
\end{figure*}

Figure \ref{fig_n11} contains the results for the $N=11$ isotones, $^{19}$O, $^{21}$Ne,
and $^{23}$Mg. In this chain the ground state of $^{19}$O is again spherical ($5/2^+$),
related to the $Z=8$ magic number. The odd neutron belongs to the  spherical
$d_{5/2}$ shell. However, both isotones $^{21}$Ne and $^{23}$Mg are $3/2^+$ prolate
states with asymptotic quantum numbers  [211]3/2. There is a $3/2^+$ excited state in
$^{19}$O at $E=96$ keV, and a $5/2^+$ excited state in $^{21}$Ne as well as in  $^{23}$Mg
at $E=351$ keV and at $E=451$ keV, respectively.

Similarly to the case of the $N=9$ chain, the comparison of the form factors in
Fig. \ref{fig_n11} shows that the shape transition produce quite different form
factors that could be easily distinguished experimentally.

\vspace{0.5cm}
{\centerline { \textbf {\textit N = 25 isotones}}}
\vspace{0.5cm}

The next example we consider is the isotonic chain $N=25$, including the spherical
nucleus $^{45}$Ca ($Z=40$), and the prolate isotones $^{47}$Ti and $^{49}$Cr. $^{45}$Ca
has a $7/2^-$ ground state that corresponds to the  $f_{7/2}$ spherical shell.
$^{47}$Ti and $^{49}$Cr have $5/2^-$ with the odd neutron in the [312]5/2 state.

A $5/2^-$ excited state is found in  $^{45}$Ca at $E= 174$ keV. Low-lying $7/2^-$
excited state are found in  $^{47}$Ti and $^{49}$Cr at $E= 159$ keV and $E=272$ keV,
respectively.

\begin{figure*}[h]
%\begin{figure}[H]
    \centering
\includegraphics[width=140mm]{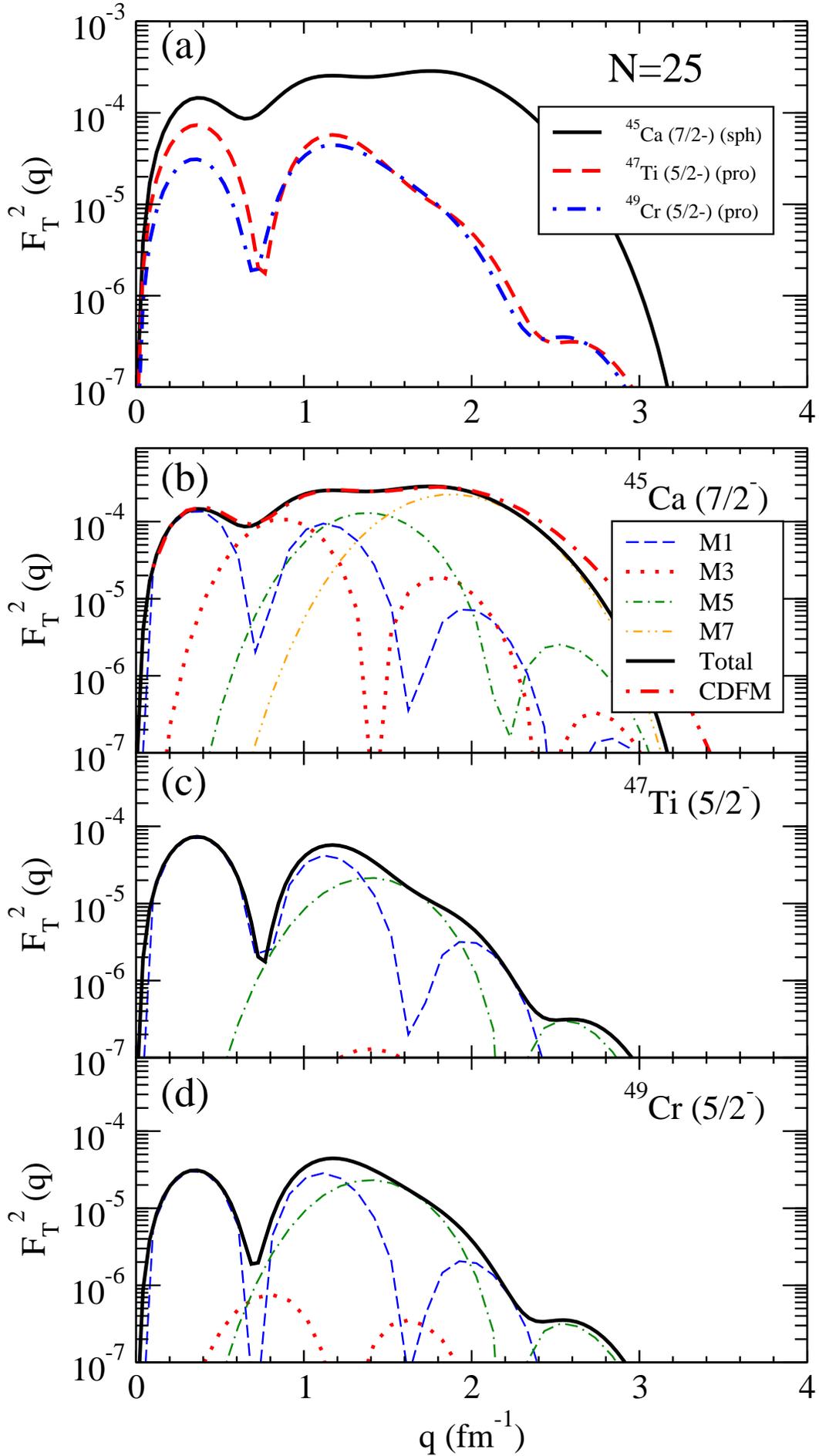}
\caption{Same as in Fig. \ref{fig_v},  but for the $N=25$ isotones  $^{45}$Ca, $^{47}$Ti, and $^{49}$Cr.
}
\label{fig_n25}
\end{figure*}

Similarly to the previous cases for $N=9$ and $N=11$, the profile of the spherical
nucleus in Fig. \ref{fig_n25} is very different from the profiles of the deformed
isotones. This is a clear signature of a structural change that will be worth
exploring experimentally.

\vspace{0.5cm}
{\centerline { \textbf {\textit N = 57 isotones}}}
\vspace{0.5cm}

Finally, we study the $N=57$ isotones, $^{99}$Mo, $^{101}$Ru, and $^{103}$Pd, where the
shape changes from spherical $1/2^+$ in $^{99}$Mo to prolate $5/2^+$ in $^{101}$Ru and
$^{103}$Pd, as can be seen in Table I. Experimentally, a $5/2^+$ excited state is
found at $E=98$ keV in  $^{99}$Mo and  $1/2^+$ excited states are found at $E=325$ keV
and $E=504$ keV in $^{101}$Ru and $^{103}$Pd, respectively. In the spherical case the
odd neutron occupies the  $3s_{1/2}$ shell, while in the prolate case the $5/2^+$
states [413]5/2 appear from the degeneracy breaking of $d_{5/2}$ and $g_{7/2}$ shells.

\begin{figure*}[h]
%\begin{figure}[H]
    \centering
\includegraphics[width=140mm]{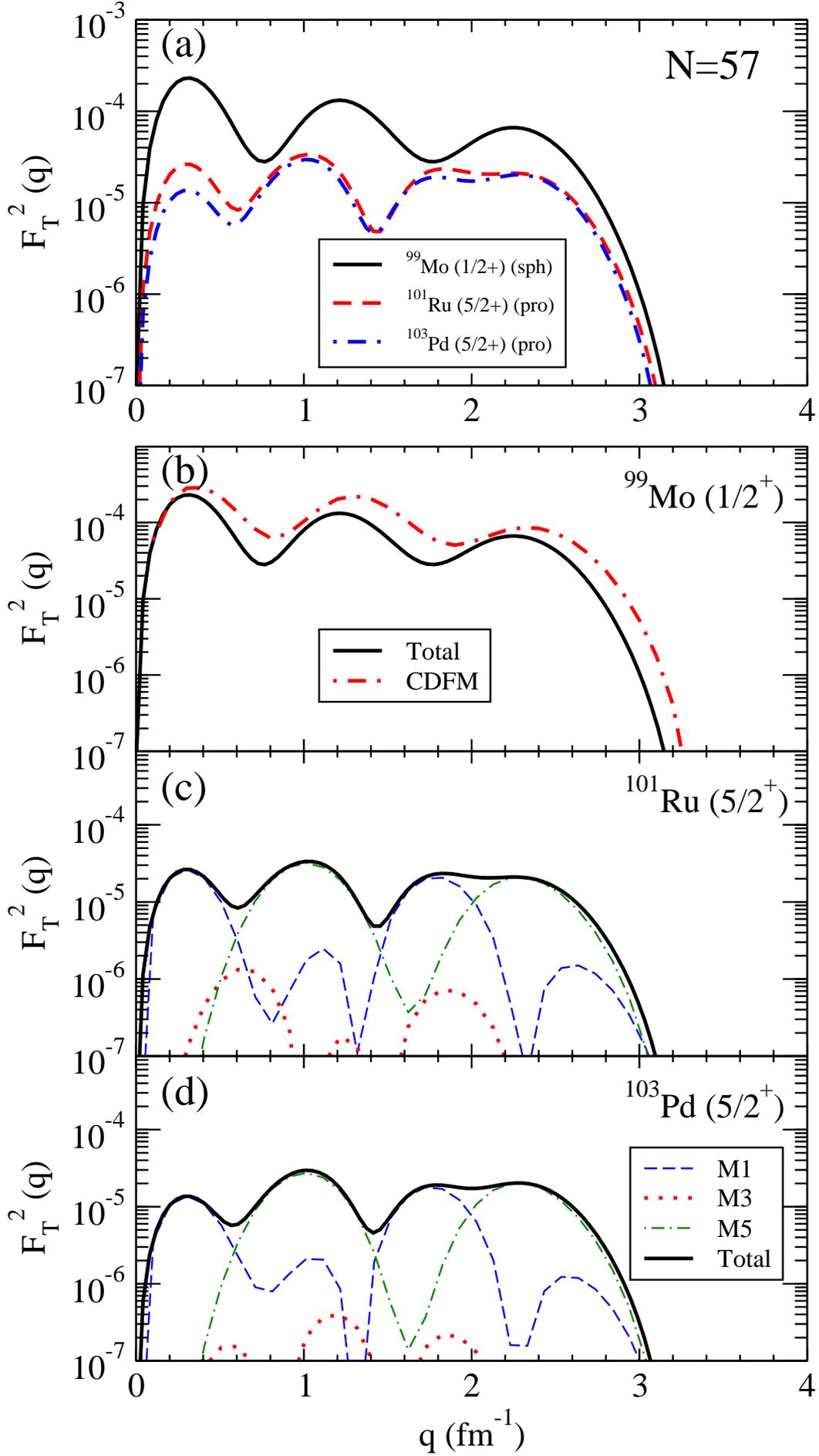}
\caption{Same as in Fig. \ref{fig_v},  but for the $N=57$ isotones $^{99}$Mo, $^{101}$Ru, and $^{103}$Pd.
}
\label{fig_n57}
\end{figure*}

The form factors in Fig. \ref{fig_n57} exhibit significant differences between the
spherical and deformed cases, although a three-peaked structure is observed in the
three isotones.

\section{Conclusions}

We have studied the sensitivity of the magnetic form factors, that could be measured
in elastic electron scattering, to structural changes of the nucleus. We focus on
odd-$Z$ and even-$N$ isotopic chains, as well as on even-$Z$ and odd-$N$ isotonic
chains, looking for changes in the experimental ground-state spins and parities within
a given chain. Nuclear structure calculations based on self-consistent deformed HF+BCS
calculations with Skyrme forces have shown that the changes of spin/parity are
related to shape transitions. 

We have considered first the chains of odd-$A$ vanadium ($^{49,51,53}$V) and aluminum
($^{25,27,29}$Al) isotopes as examples of spherical and deformed nuclei where the
ground-state spin and parity is the same in all the members of the chain. This is
the expected behavior when the nuclear structure in general and the odd-proton wave
function in particular does not change significantly with the addition of an even
number of neutrons.
We have shown that the magnetic form factors of the isotopes within these chains
are basically the same. After these two examples we studied isotopic chains of
fluorine ($^{17,19,21}$F), manganese ($^{51,53,55}$Mn), iodine ($^{125,127,129}$I), and
cesium ($^{131,133,135}$Cs), as well as isotonic chains with $N=9$ ($^{15}$C, $^{17}$O,
$^{19}$Ne),  $N=11$ ($^{19}$O, $^{21}$Ne, $^{23}$Mg), $N=25$ ($^{45}$Ca, $^{47}$Ti,
$^{49}$Cr), and $N=57$ ($^{99}$Mo, $^{101}$Ru, $^{103}$Pd). In all of these chains the
ground-state spins and parities change within the chain. The corresponding nuclear
structure calculations demonstrate the correlation between a change of the
ground-state spin and parity and a shape transition within the chain. The results of
the calculations of the elastic magnetic form factors exhibit quite different profiles
for different deformations that could be interpreted as signatures of shape transitions.

Comparison with experiment in the few stable nuclei where this information is available
(see also \cite{merino}) demonstrates the ability of the method to reproduce the data,
especially at low momentum transfer and even at higher values beyond 2 fm$^{-1}$ when
short-range correlations are included. These nucleon-nucleon correlations, taken into
account by the CDFM method, shift the tails of the form factors at higher momentum
transfer. This comparison is also needed to trust the predictions for unstable nuclei,
where no data are available yet.

It would be very interesting to check experimentally the predictions for the magnetic
form factors for the  isotopic and isotonic chains analyzed here, which have been
found to show a shape transition. The present theoretical study is timely, as new
experimental electron-scattering facilities are expected to deal soon with unstable
nuclei.

\begin{acknowledgments}
  This work was supported by Ministerio de Ciencia e Innovaci\'on  MCI/AEI/FEDER,UE
  (Spain) under Contract Nos. PGC2018-093636-B-I00 and RTI2018-098868-B-I00.
  Two of the authors (D.N.K. and A.N.A.) are grateful for support of the Bulgarian
  Science Fund under Contract No. KP-06-N38/1.

\end{acknowledgments}

\begin{table*}[th]
  \caption{Experimental spin/parity ($I^{\pi}$) and half-life ($T_{1/2}$), calculated and
    measured charge root-mean-square radii ($r_{c,{\rm th}}$ and $r_{c,{\rm exp}}$ \cite{angeli}
    in fm), calculated quadrupole deformation ($\beta_p$), calculated and measured
    spectroscopic electric quadrupole moment ($Q_{\rm lab,th}$ and $Q_{\rm lab,exp}$ \cite{stone}
    in barns), and calculated (in spherical limit and deformed) and measured magnetic
    moments ($ \mu_{\rm th, sph} , \mu_{\rm th, def} , \mu_{\rm exp}$  \cite{stone} in  $[\mu_N ]$)
    in isotopic chains with $Z=$ 9, 13, 23, 25, 53, and 55. } 
{\begin{tabular}{ccccccccccc} \hline \hline \\
    Nucleus   & $I^{\pi}$ & $T_{1/2}$ & $r_{c,{\rm th}}$  &  $r_{c,{\rm exp}}$  &
    $\beta_p$  & $Q_{\rm lab, th}$ & $Q_{\rm lab, exp}$  & $\mu_{\rm th, sph} $ &
    $\mu_{\rm th, def} $ & $\mu_{\rm exp} $  \\ \\

$^{17}$F  & $5/2^+$ &      64.49 s       & 2.8497 &   -        &   sph    &   sph    &    0.058(4)   & $+4.793$ &   -      & $+4.7213(3)$   \\ 
$^{19}$F  & $1/2^+$ &      stable        & 2.8591 & 2.8976(25) & $+0.193$ &     0    &       -       & $+4.793$ & $+2.275$ & $+2.628868(8)$  \\
$^{21}$F  & $5/2^+$ &      4.158 s       & 2.8460 &    -       & $-0.121$ & $-0.038$ &       -       & $+4.793$ &   3.730  &   3.93(5)        \\ \\

$^{25}$Al & $5/2^+$ &      7.183 s       & 3.1317 &     -      & $+0.303$ & $+0.166$ &      -        & $+4.793$ & $+3.795$ &   3.6455(12)     \\
$^{27}$Al & $5/2^+$ &      stable        & 3.0948 & 3.0610(31) & $+0.200$ & $+0.107$ & $+0.1466(10)$ & $+4.793$ & $+3.768$ & $+3.6415069(7)$   \\
$^{29}$Al & $5/2^+$ &      6.56 min      & 3.0934 &    -       & $+0.128$ & $+0.069$ &     -         & $+4.793$ & $+3.743$ &         -          \\ \\

$^{49}$V  & $7/2^-$ &      330 d         & 3.6201 &     -      &    sph   &    sph   &     -         & $+5.793$ &     -    &   4.47(5)       \\
$^{51}$V  & $7/2^-$ &      stable        & 3.6222 & 3.6002(22) &    sph   &    sph   & $-0.043(5)$   & $+5.793$ &     -    & $+5.1487057(2)$   \\
$^{53}$V  & $7/2^-$ &     1.543 min      & 3.6388 &     -      &    sph   &    sph   &     -         & $+5.793$ &     -    &         -          \\ \\

$^{51}$Mn & $5/2^-$ &      46.2 min      & 3.7235 & 3.7026(212)& $+0.244$ & $+0.367$ &    -          & $+5.793$ & $+3.591$ &   3.5683(13)    \\
$^{53}$Mn & $7/2^-$ & 3.7$\times 10^6$ y & 3.6897 & 3.6662(76) &    sph   &  sph     &    -          & $+5.793$ &    -     &   5.024(7)       \\
$^{55}$Mn & $5/2^-$ &      stable        & 3.7316 & 3.7057(22) & $+0.212$ & $+0.321$ & $+0.33(1)$    & $+5.793$ & $+3.532$ & $+3.46871790(9)$   \\ \\

$^{125}$I  & $5/2^+$ &     59.4 d         & 4.7507 &     -      & $-0.148$ & $-0.779$ & $-0.776(17)$ & $+1.717$ & $+3.219$ &  2.821(5)      \\
$^{127}$I  & $5/2^+$ &     stable         & 4.7596 & 4.7500(81) & $-0.137$ & $-0.725$ & $-0.710(10)$ & $+1.717$ & $+3.259$ & $+2.81327(8)$   \\
$^{129}$I  & $7/2^+$ & 1.6$\times 10^7$ y & 4.7545 &     -      &    sph   &  sph     & $-0.498(7)$  & $+1.717$ &   -      & $+2.6210(3)$     \\ \\

$^{131}$Cs & $5/2^+$ &     9.689 d        & 4.8085 & 4.8026(47) & $-0.134$ & $-0.751$ & $-0.575(6)$   & $+1.717$ & $+2.043$ & $+3.53(2)$      \\    
$^{133}$Cs & $7/2^+$ &     stable         & 4.8069 & 4.8041(46) &    sph   &  sph     & $-0.00355(4)$ & $+1.717$ &     -    & $+2.58205(3)$    \\
$^{135}$Cs & $7/2^+$ & 2.3$\times 10^6$ y & 4.8107 & 4.8067(47) &    sph   &  sph     & $+0.050(2)$   & $+1.717$ &    -     & $+2.7324(2)$      \\ \\

\hline \hline
\label{table_isotopes}
\end{tabular}}
\end{table*}

\begin{table*}[th]
  \caption{Same as in Table \ref{table_isotopes}, but for isotonic chains with $N=$ 9, 11,
    25, and 57.} 
{\begin{tabular}{ccccccccccc} \hline \hline \\
    Nucleus   & $I^{\pi}$ & $T_{1/2}$ & $r_{c,{\rm th}}$  &  $r_{c,{\rm exp}}$ \cite{angeli} &
    $\beta_p$  & $Q_{\rm lab, th}$ & $Q_{\rm lab, exp}$ \cite{stone} & $\mu_{\rm th, sph} $ &
    $\mu_{\rm th, def} $ & $\mu_{\rm exp} $ \cite{stone}  \\ \\

$^{15}$C   & $1/2^+$ & 2.449 s  & 2.5764 &     -       & $+0.122$ &    0     &     -       & $-1.913$ & $-1.361$ &   1.720(9 )   \\
$^{17}$O   & $5/2^+$ & stable   & 2.7476 & 2.6932(75)  &   sph    &    sph   & $-0.02578$  & $-1.913$ &   -      & $-1.89379(9)$  \\
$^{19}$Ne  & $1/2^+$ & 17.22 s  & 2.9543 & 3.0082(40)  & $+0.189$ &    0     &     -       & $-1.913$ & $-1.521$ & $-1.88542(8)$    \\ \\

$^{19}$O   & $5/2^+$ & 26.88 s  & 2.7445 &     -       &   sph    &    sph   &  0.0037(4)  & $-1.913$ &   -      &  1.53195(7)   \\
$^{21}$Ne  & $3/2^+$ & stable   & 2.9582 & 2.9695(33)  & $+0.276$ & $+0.058$ & $+0.103(8)$ & $-1.913$ & $-0.718$ & $-0.661797(5)$ \\
$^{23}$Mg  & $3/2^+$ & 11.317 s & 3.1136 &     -       & $+0.398$ & $+0.112$ &  0.125(5)   & $-1.913$ & $-0.732$ &  0.5364(3)      \\ \\

$^{45}$Ca  & $7/2^-$ & 162.61 d & 3.4973 & 3.4944(21)  &   sph    &  sph     & $+0.046(14)$ & $-1.913$ &   -      & $-1.3274(14)$  \\
$^{47}$Ti  & $5/2^-$ & stable   & 3.6072 & 3.5962(19)  & $+0.185$ & $+0.230$ & $+0.30(2)$   & $-1.913$ & $-0.816$ & $-0.78848(1)$   \\
$^{49}$Cr  & $5/2^-$ & 42.3 min & 3.6932 &     -       & $+0.260$ & $+0.369$ &      -       & $-1.913$ & $-0.841$ &  0.476(3)        \\ \\

$^{99}$Mo  & $1/2^+$ & 65.976 h & 4.3769 &     -       &    sph   &    0     &    -        &  $+1.488$ &   -      &  0.375(3)   \\
$^{101}$Ru & $5/2^+$ & stable   & 4.4486 & 4.4606(20)  & $+0.148$ & $+0.565$ & $+0.46(2)$  &  $+1.488$ & $-0.921$ & $-0.719(6)$  \\
$^{103}$Pd & $5/2^+$ & 16.991 d & 4.4936 &     -       & $+0.161$ & $+0.656$ &    -        &  $+1.488$ & $-0.799$ &     -         \\ \\

\hline \hline
\label{table_isotones}
\end{tabular}}
\end{table*}

%\newpage
\clearpage

%%%%%%%%%%%%%%%%%%%%%%%%%%%%Fig1%%%%%%%%%%%%%%%%%%%%%%%%%%%%%%%%%%%%%%%%%%%%%%%%%%%%%%%%
%\end{figure}
%%%%%%%%%%%%%%%%%%%%%%%%%%%%%%%%%%%%%%%%%%%%%%%%%%%%%%%%%%%%%%%%%%%%%%%%%%%%%%%%%%%%%%%%%

%%%%%%%%%%%%%%%%%%%%%%%%%%%%Fig2%%%%%%%%%%%%%%%%%%%%%%%%%%%%%%%%%%%%%%%%%%%%%%%%%%%%%%%%
%\end{figure}
%%%%%%%%%%%%%%%%%%%%%%%%%%%%%%%%%%%%%%%%%%%%%%%%%%%%%%%%%%%%%%%%%%%%%%%%%%%%%%%%%%%%%%%%%

%%%%%%%%%%%%%%%%%%%%%%%%%%%%Fig3%%%%%%%%%%%%%%%%%%%%%%%%%%%%%%%%%%%%%%%%%%%%%%%%%%%%%%%%
%\end{figure}
%%%%%%%%%%%%%%%%%%%%%%%%%%%%%%%%%%%%%%%%%%%%%%%%%%%%%%%%%%%%%%%%%%%%%%%%%%%%%%%%%%%%%%%%%

%%%%%%%%%%%%%%%%%%%%%%%%%%%%Fig4%%%%%%%%%%%%%%%%%%%%%%%%%%%%%%%%%%%%%%%%%%%%%%%%%%%%%%%%
%\end{figure}
%%%%%%%%%%%%%%%%%%%%%%%%%%%%%%%%%%%%%%%%%%%%%%%%%%%%%%%%%%%%%%%%%%%%%%%%%%%%%%%%%%%%%%%%%

%%%%%%%%%%%%%%%%%%%%%%%%%%%%Fig5%%%%%%%%%%%%%%%%%%%%%%%%%%%%%%%%%%%%%%%%%%%%%%%%%%%%%%%%
%\end{figure}
%%%%%%%%%%%%%%%%%%%%%%%%%%%%%%%%%%%%%%%%%%%%%%%%%%%%%%%%%%%%%%%%%%%%%%%%%%%%%%%%%%%%%%%%%

%%%%%%%%%%%%%%%%%%%%%%%%%%%%Fig6%%%%%%%%%%%%%%%%%%%%%%%%%%%%%%%%%%%%%%%%%%%%%%%%%%%%%%%%
%\end{figure}
%%%%%%%%%%%%%%%%%%%%%%%%%%%%%%%%%%%%%%%%%%%%%%%%%%%%%%%%%%%%%%%%%%%%%%%%%%%%%%%%%%%%%%%%%

%%%%%%%%%%%%%%%%%%%%%%%%%%%%Fig7%%%%%%%%%%%%%%%%%%%%%%%%%%%%%%%%%%%%%%%%%%%%%%%%%%%%%%%%
%\end{figure}
%%%%%%%%%%%%%%%%%%%%%%%%%%%%%%%%%%%%%%%%%%%%%%%%%%%%%%%%%%%%%%%%%%%%%%%%%%%%%%%%%%%%%%%%%

%%%%%%%%%%%%%%%%%%%%%%%%%%%%Fig8%%%%%%%%%%%%%%%%%%%%%%%%%%%%%%%%%%%%%%%%%%%%%%%%%%%%%%%%
%\end{figure}
%%%%%%%%%%%%%%%%%%%%%%%%%%%%%%%%%%%%%%%%%%%%%%%%%%%%%%%%%%%%%%%%%%%%%%%%%%%%%%%%%%%%%%%%%

%%%%%%%%%%%%%%%%%%%%%%%%%%%%Fig9%%%%%%%%%%%%%%%%%%%%%%%%%%%%%%%%%%%%%%%%%%%%%%%%%%%%%%%%
%\end{figure}
%%%%%%%%%%%%%%%%%%%%%%%%%%%%%%%%%%%%%%%%%%%%%%%%%%%%%%%%%%%%%%%%%%%%%%%%%%%%%%%%%%%%%%%%%

%%%%%%%%%%%%%%%%%%%%%%%%%%%%Fig10%%%%%%%%%%%%%%%%%%%%%%%%%%%%%%%%%%%%%%%%%%%%%%%%%%%%%%%%
%\end{figure}
%%%%%%%%%%%%%%%%%%%%%%%%%%%%%%%%%%%%%%%%%%%%%%%%%%%%%%%%%%%%%%%%%%%%%%%%%%%%%%%%%%%%%%%%%

\end{document}